\documentclass[namedreferences]{solarphysics}

\usepackage[hyperref,optionalrh,showbiblabels]{spr-sola-addons} % For Solar Physics 
\usepackage{graphicx}        % For eps figures, newer & more powerfull
\usepackage{color}           % For color text: \color command
\usepackage{breakurl}        % For breaking URLs easily trough lines
\usepackage{multirow}
\renewcommand{\vec}[1]{{\mathbfit #1}}

\newcommand{\aap}{    {\it Astron. Astrophys.}}

\newcommand{\apj}{    {\it Astrophys. J.}}
\newcommand{\apjl}{   {\it Astrophys. J. Lett.}}
\newcommand{\apjs}{   {\it Astrophys. J. Suppl. Series}}

\newcommand{\solphys}{{\it Solar Phys.}}
 
\newcommand{\ssr}{    {\it Space Sci. Rev.}} 
\chardef\us=`\_

\begin{document}

\begin{article}
\begin{opening}

\title{Oscillations in the Flaring Active Region NOAA 11272\\ {\it Solar Physics}}

\author[addressref={aff1},corref,email={sandra.cuellar@inpe.br}]{\inits{S.M.}\fnm{S.M.}~\lnm{Conde Cuellar}}%\sep
\author[addressref=aff1,email={joaquim.costa@inpe.br}]{\inits{J.E.R.}\fnm{J.E.R.}~\lnm{Costa}}%\sep
\author[addressref=aff1,email={eduardo.montana@inpe.br}]{\inits{C.E.}\fnm{C.E.}~\lnm{Cede\~{n}o Monta\~{n}a}}%\sep

\address[id=aff1]{Instituto Nacional de Pesquisas Espaciais-INPE, Av. dos Astronautas, 1758, S{\~a}o Jos{\'e} dos Campos, 12227-010 SP, Brazil}

\runningauthor{S.M. Conde Cuellar et al.}
\runningtitle{Oscillations in the Flaring Active Region NOAA 11272}

\begin{abstract}

We studied waves seen during the class C1.9 flare that occurred in Active Region NOAA 11272 on SOL2011-08-17. We found standing waves with periods in the 9- and 19-minute band in six extreme ultraviolet (EUV) wavelengths of the SDO/AIA instrument. We succeeded in identifying the magnetic arc where the flare started and two neighbour loops that were disturbed in sequence. The analysed standing waves spatially coincide with these observed EUV loops. To study the wave characteristics along the loops, we extrapolated field lines from the line-of-sight magnetograms using the force-free approximation in the linear regime. We used atmosphere models to determine the mass density and temperature at each height of the loop. Then, we calculated the sound and Alfv{\'e}n speeds using densities $10^8 \lesssim n_i \lesssim 10^{17}$ cm$^{-3}$ and temperatures $10^3 \lesssim T \lesssim 10^7$ K. The brightness asymmetry in the observed standing waves resembles the Alfv{\'e}n speed distribution along the loops, but the atmospheric model we used needs higher densities to explain the observed periods. 

\end{abstract}


\end{opening}
%-------------------------------------------------

\section{Introduction}
     \label{S-Introduction} 

Magnetohydrodynamic (MHD) waves in the solar corona have been studied for many years \citep{LS82,A04,MJE13}. Particularly, waves in coronal loops have been recently analysed in the quiescent and transient Sun \citep{D06,KNS12}. These waves have been classified, depending on their period, phase,  amplitude, phase speed, etc. as kink, sausage, magnetoacoustic fast, or slow waves, among others \citep{R81a,R81b}.

In the solar transient regime it is common to see flaring events, and it is interesting to analyse waves associated with them, in some cases in a recursive mode that is called quasi periodic pulsations (QPPs) \citep{NM09}. They can be observed at all stages of the flare, in different bands at the same time, and can have periods from seconds to several minutes.

The \textit{Atmospheric Imaging Assembly} (AIA) on board the \textit{Solar Dynamics Observatory} (SDO) provides images in ultraviolet (UV) and extreme ultraviolet (EUV) with a pixel size of 0.6 arcsec and a temporal cadence of 12 seconds \citep{LAB12}. This allows us to observe waves in coronal loops during transient events \citep{NNV13}. On the other hand, with the \textit{Helioseismic and Magnetic Imager} (HMI) instrument on board the SDO, it is possible to obtain magnetograms in the line-of-sight (LOS) every 45 seconds with a pixel size of 0.5 arcsec \citep{SSBK12}. 

The waves detected are commonly analysed using wavelet transform to determine the period, power, amplitude, and phase  \citep{DH00,DMH04}. One method used is named pixelised wavelet filtering (PWF). It applies the continuous wavelet transform pixel by pixel with a Morlet mother function \citep{TC98} in the temporal series of a 3D data cube. This method allows us to determine periods and to spatially and temporally analyse waves that are found through the narrowband dynamic maps \citep{SN08,SNAO10}.

Different studies of the oscillatory process along coronal loops were made using numerical models in 2D \citep{MES04,MHE10,EM04}. We present a 3D study of the velocity profile for coronal loops. To do this, we detected waves on images that were obtained with SDO/AIA, using the PWF method. After this, we extrapolated the magnetic field into the corona using the linear force-free approximation \citep{NR72,SSC08} from the magnetograms obtained with SDO/HMI. Thus, we reproduced the field lines that best match the coronal loops observed in EUV to study the variation of physical parameters along them. To determine these parameters, we used atmospheric models for flares \citep{MAVN80} and the quiet Sun \citep{AL08}.

To present our results, we describe the data cube we analysed and the techniques we used for data processing in Section \ref{S-aug}. The waves found in the EUV images are shown in Section \ref{S-general} and the calculated speed of the waves along the extrapolated field lines in Section \ref{S-features}. Finally, we provide a brief summary and final remarks in Section \ref{S-otro}.

\section{Observations and Data Processing}
\label{S-aug}

A class C1.9 flare was recorded with GOES-15 spacecraft on 17 August 2011 from 01:44 UT to 03:44 UT.  This flare occurred in Active Region NOAA 11272, which was located close to the south-east limb on the solar disk, at S19E53 ($-717''$,$-372''$).  We analysed this region during two hours and thirty minutes, from 01:15:00 UT (before the flaring activity) until 03:44:48 UT. 

\begin{figure}[h!]
\centerline{%
 \includegraphics[width=0.5\textwidth,clip=]{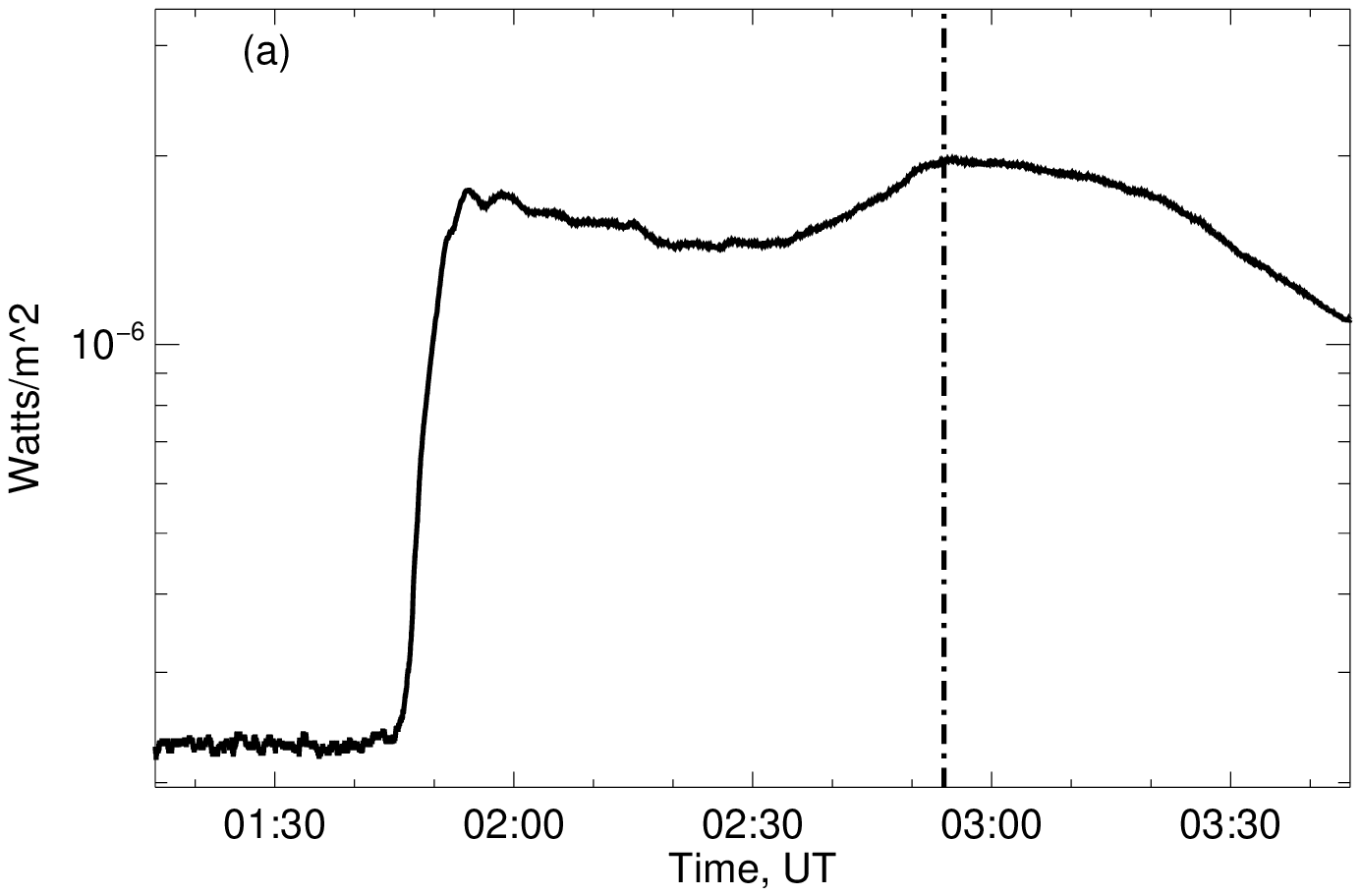}%
 \includegraphics[width=0.5\textwidth,clip=]{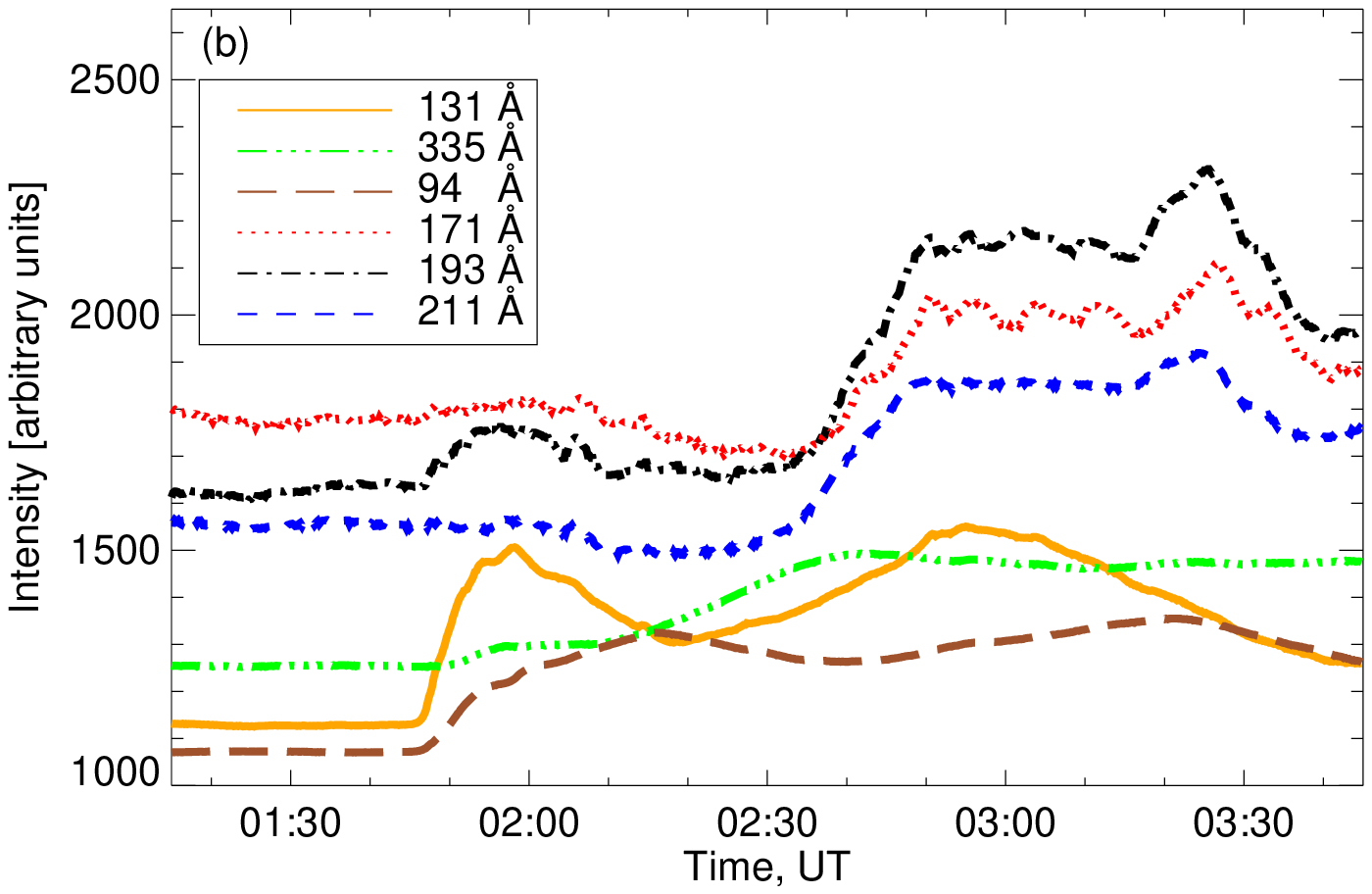} %
 }%
\centerline{%
 \includegraphics[width=0.5\textwidth,clip=]{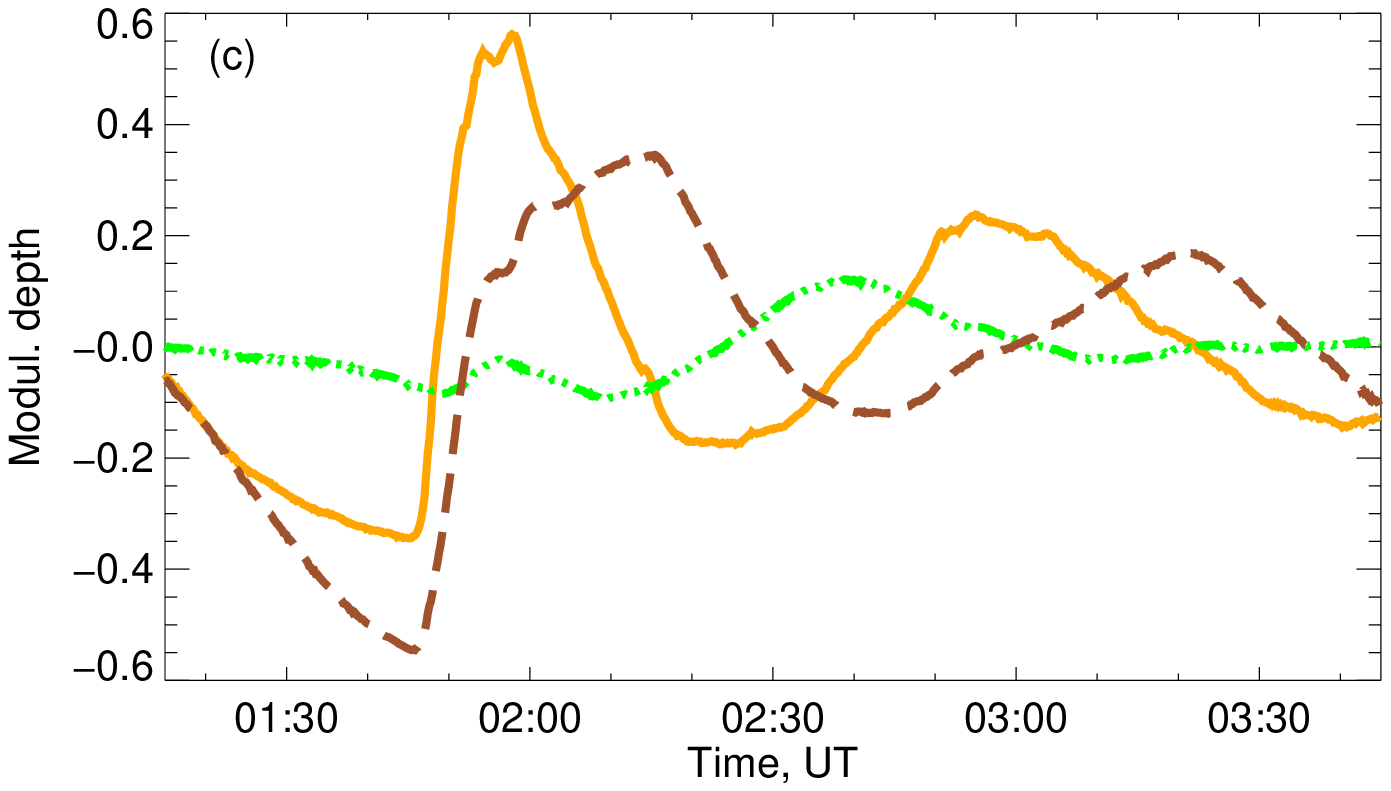} %
 \includegraphics[width=0.5\textwidth,clip=]{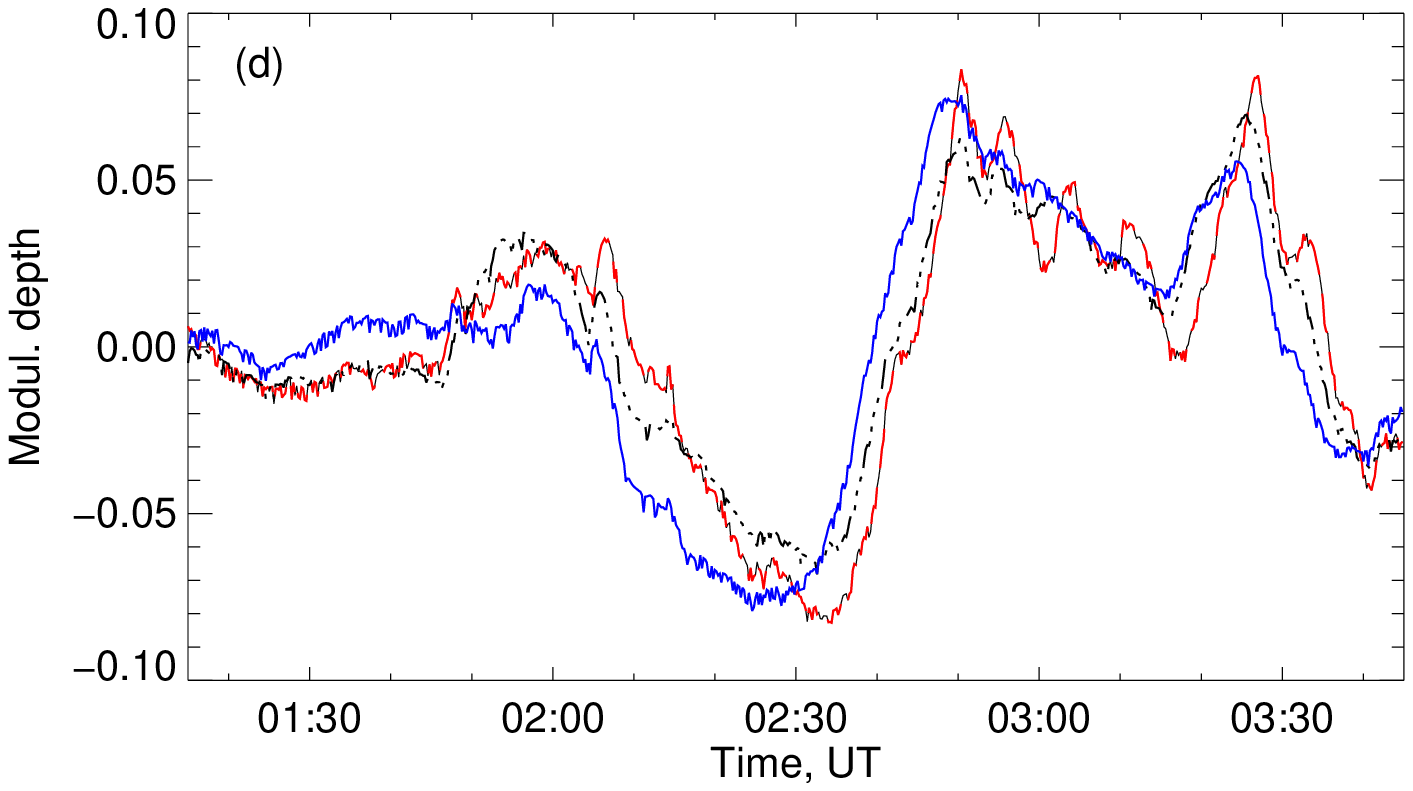} %
 }%
\caption{(a) Light curve from GOES-15 during the C1.9 flare. The  vertical dash-dotted line indicates the moment when the maximum occurred at 02:54 UT. (b) The time profile of the intensity on the ROI during the flare from six SDO/AIA bandpasses. Here, the beginning and the maximum moments of the flare are shown. Emission modulation depth at (c) 131, 335, and 94 \r{A} and (d) 171, 193, and 211 \r{A} using the same colours and line style as in (b). }
\label{F-goes}
\end{figure}

Figure \ref{F-goes}(a) shows the X-ray flux 1.0$-$8.0 \r{A} from the GOES-15 spacecraft during this flare. The  vertical dash-dotted line indicates the moment when the maximum occurred at 02:54 UT. The light curve rises after 01:44 UT, reaching its maximum, and then it decreases.

The flare was observed with the SDO/AIA\footnote{The images are courtesy of \textit{Solar Dynamics Observatory}, which is part of the NASA ``Living with a star'' program.} in six EUV wavelengths, 131, 171, 193, 211, 335, and 94 \r{A}. The images were obtained from level-1 calibrated data with temporal cadence of 12 seconds. These datasets were processed and co-aligned using the SolarSoft routine {\it aia{\_}prep.pro}. For each of six EUV bandpasses, we generated four 3D data cubes (two spatial components and time) with 300 frames each, \textit{i.e.} one hour of observation for each cube. We made two principal cubes in the interval of 01:45:00 UT and 03:44:48 UT that correspond to the whole flare period, and two other cubes that overlap by thirty minutes, \textit{i.e.} from 01:15:00 UT  to 03:14:48 UT.  We did not use images at 4500 \r{A} because of its low temporal cadence of 3600 seconds (one hour) or at 1700, 1600, and 304 \r{A} because the loops of our interest were not detected in these bandpasses.

\begin{figure}[h!]
\centerline{%
 \includegraphics[width=0.85\textwidth,clip=]{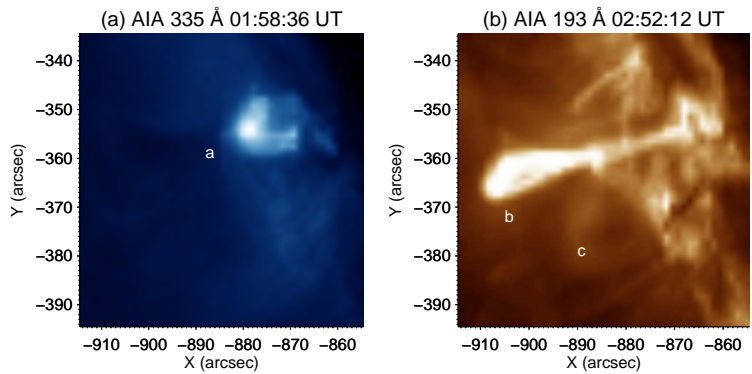}%
 }%
\caption{(a) Loops \textbf{a} where the C1.9 flare started in AIA 335 \r{A}. (b) Neighbouring loops \textbf{b} and \textbf{c} perturbed by the flare observed in 193 \r{A}.}
\label{F-rois}
\end{figure}

On the other hand, we used two LOS magnetograms obtained from SDO/HMI at 02:00:39 and 02:42:39 UT. The data, level 1.5, were processed and aligned with the AIA images using the {\it aia{\_}prep.pro} routine. We extrapolated the magnetic field lines for these magnetograms and compared them with loops  \textbf{a},  \textbf{b}, and  \textbf{c} seen in the AIA images (see Figure \ref{F-rois}) in similar observation time.

The region of interest (ROI) selected for our study was chosen with dimensions $66'' \times 66''$. Figure \ref{F-rois} shows this ROI and two notable moments during the flare. Loop \textbf{a} represents the place where the flaring activity began (Figure \ref{F-rois}(a)). Loops \textbf{b} and \textbf{c} were perturbed during the flare (Figure \ref{F-rois}(b)). 

The time profile of the intensity (in arbitrary units) shown in Figure \ref{F-goes}(b)  corresponds to the temporal signal integrated over this ROI.  For 131, 335, and 94 \r{A} the intensity is re-scaled through multiplying by an \textit{ad hoc} factor of 2.5 and adding $10^3$.

In Figure \ref{F-goes}(c) we show the time profiles of the modulation depth for 131, 335, and 94 \r{A} and in Figure \ref{F-goes}(d) we show the profiles for 171, 193, and 211 \r{A} using the same colour and line style as in Figure \ref{F-goes}(b).

The analysed period of this flare shows two peaks in the GOES-15 time profile, but in the EUV the modulations seem to agree with a QPPs regime as described in \cite{NM09}. The modulation depth was calculated using the equation $\Delta I / I=(I(t)-I_0)/I_0$, where $I(t)$ is the intensity and $I_0$ is the time profile smoothed over an interval of 75 minutes  \citep{KMNS10}. The highest modulations were found for 131, 94, and 335 \r{A}, around of 60\%, 40\%, and 10\%, respectively. For 171, 193, and 211 \r{A} we found a signal modulated by 8\%. 

We used the pixelised wavelet filtering (PWF) method \citep{SN08,SNAO10} to determine the presence of oscillations. With this, we studied the spatial distribution and temporal evolution of the oscillation sources in the coronal loops shown in Figure \ref{F-rois}.  For the temporal signal of each pixel, we applied the continuous wavelet transform inside the 3D data cube of the selected ROI using a Morlet mother function \citep{TC98}\footnote{We used the software provided by C. Torrence and G. P. Compo for the wavelet analysis, which is available at http://paos.colorado.edu/research/wavelets/.}. Then, we filtered this by the period-bands that were previously chosen and calculated the inverse wavelet transform on the resulting 4D data cube (two spatial components, time and period).  These period-bands were selected in the range of 0.4$-$30 minutes.  We chose each period-band from $dt\times2^{n-1}$ to $dt\times2^n$, where $dt$=24 s and $n$ is varying between 1 and 6. The periods of 9 and 19 minutes had widths of 6 and 12 minutes, respectively.  As a result, we obtained the temporal signal for a specific period-band, \textit{i.e.} a narrowband map that enabled us to see the spatial distribution of the amplitude in the period-band of our interest.  Finally, to study the temporal evolution of the oscillations we found, we constructed dynamical narrowband maps, \textit{i.e.} videos.

\section{Waves Detected} 
      \label{S-general}      

For loops \textbf{a} and \textbf{b} of Figure \ref{F-rois}, we found oscillations with periods on the 9-minute band ($9\pm3$ min). They were detected in the EUV wavelengths except in 94 and 131 \r{A} for loop \textbf{b}. During approximately forty minutes in 171 and 193 \r{A} and in one hour for the other bandpasses, the oscillatory presence was detected in the loops.  Dynamical narrowband maps showed standing waves from 01:30:00 UT to 02:29:48 UT for loop \textbf{a} and between 02:15:00 UT and 03:14:48 UT for loop \textbf{b}. \\
Two videos that are available on-line show the standing wave along loop {\textbf{a}} in 335 \r{A} and two others at loop {\textbf{b}} in 211 \r{A}.

\begin{figure}[h!]
\centerline{%
 \includegraphics[width=0.9\textwidth,clip=]{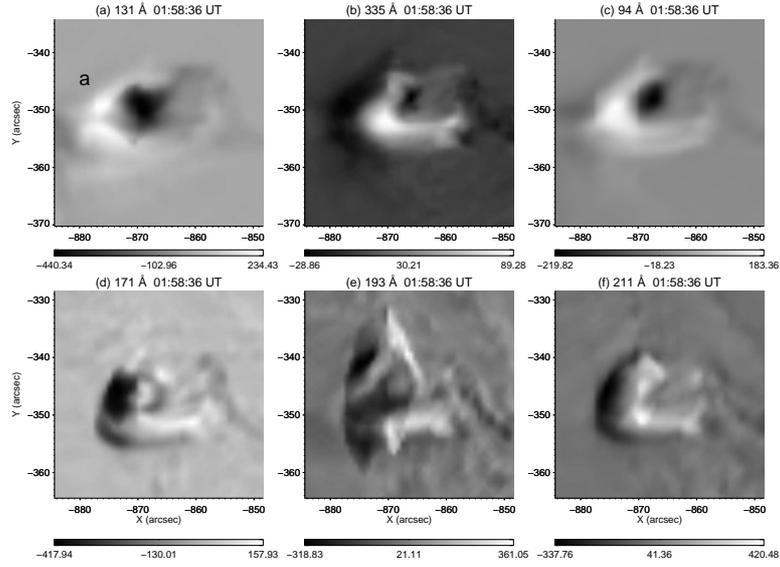}%
 }%
\caption{Snapshots of narrowband maps for amplitude (arbitrary units) of the 19-minute waves in loop \textbf{a}. The waves were seen in 131 \r{A} (a), 335 \r{A} (b), 94 \r{A} (c), 171 \r{A} (d), 193 \r{A} (e), and 211 \r{A} (f). The colour bars indicate the variation between the maximum and minimum amplitudes. The on-line video shows the standing wave for 171 \r{A}. }
\label{F-mdarc}
\end{figure}

Similarly, we observed oscillations in the 19-minute band ($19\pm6$ min) at loops \textbf{a}, \textbf{b}, and \textbf{c} shown in Figure \ref{F-rois} in the six EUV bandpasses, during approximately one hour.  Figures \ref{F-mdarc}(a)$-$(f) show snapshots of narrowband maps for the amplitude. They were observed from 01:15:00 UT (before the flare start time registered in the GOES curve) to 02:44:48 UT.  The waveform seen at 01:58:36 UT coincides with the shape of loop \textbf{a} (see Figure \ref{F-rois}(a)). Comparing the maximum (bright details) and minimum (dark) of amplitude, we see the waves in phase between 131, 335, and 94 \r{A} (Figures  \ref{F-mdarc}(a)$-$(c)).  However, between 171, 193, and 211 \r{A} (Figures  \ref{F-mdarc}(d)$-$(f)) the two groups are out of phase.

\begin{figure}[h!] 
\centerline{%
 \includegraphics[width=0.9\textwidth,clip=]{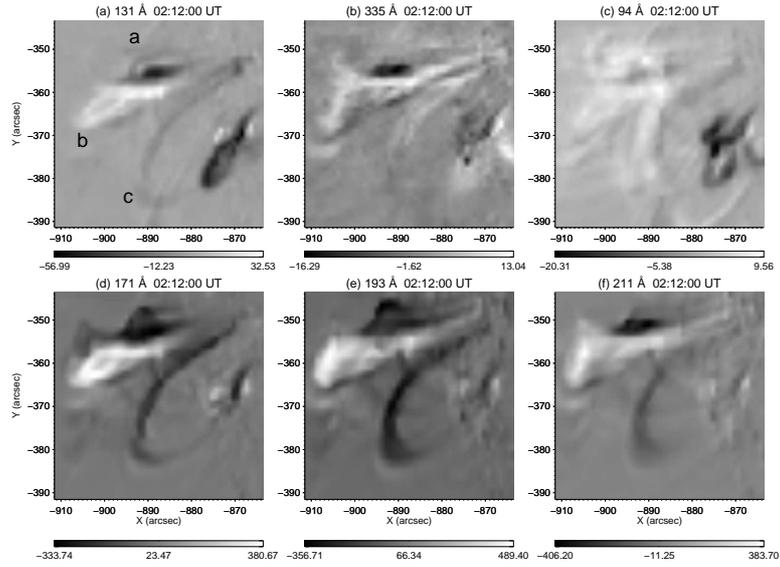}%
 }%
\caption{Snapshots of narrowband maps for the amplitude (arbitrary units) of 19-minute waves in loops \textbf{b} and \textbf{c}. The waves were detected in 131 \r{A} (a), 335 \r{A} (b), 94 \r{A} (c), 171 \r{A} (d), 193 \r{A} (e), and 211 \r{A} (f). The colour bars indicate the variation between the maximum and minimum amplitudes.  The on-line video shows the standing wave for 171 \r{A}. }
\label{F-mdrb}
\end{figure}

Furthermore, loops \textbf{b} and \textbf{c} were perturbed after starting the flare (Figure \ref{F-mdrb}). The waveform seen at 02:12:00 UT reproduces the form of the loops shown in Figure \ref{F-rois}(b). In particular, loop \textbf{c} is highlighted in the snapshots of 171, 193, and 211 \r{A}. In loop \textbf{a}, the oscillations were detected between 02:15:00 UT and 03:44:48 UT. In loops \textbf{b} and \textbf{c} they appeared from 01:45:00 UT to 02:44:48 UT and were clearly visible in the 171, 193, and 211 \r{A} bandpasses (Figures \ref{F-mdrb}(d)$-$(f)). The variation between maximum and minimum amplitudes seen in the narrowband maps shows waves in phase for loops \textbf{b} and \textbf{c} in all EUV bandpasses.


We found a high spatial coincidence between the appearance of the waveform shown in Figures \ref{F-mdarc} and \ref{F-mdrb} and the shape of loops in Figure \ref{F-rois} during the entire time of the analysis. Furthermore, the waves appear in all moments of the flare, starting at loop \textbf{a} and continuing in loops \textbf{b} and \textbf{c}. This was confirmed through the dynamical narrowband maps for the amplitude and phase in all six EUV bandpasses.

\begin{figure}[h!]
\centerline{%
\includegraphics[width=1.\textwidth,clip=]{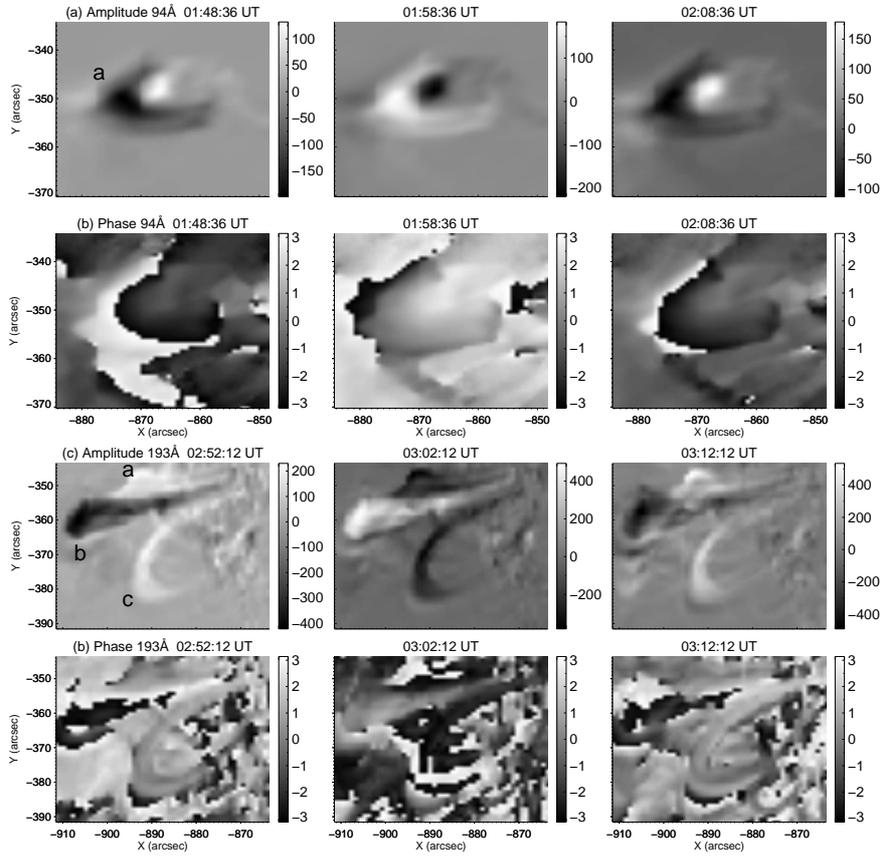}%
}%
\caption{Snapshots of amplitude (arbitrary units) and phase (rad) narrowband maps in \hspace{1cm}94 \r{A} (a), (b), and 193 \r{A} (c), (d) for 19-minute standing waves.}
\label{F-diag}
\end{figure}

Figures \ref{F-diag}(a) and \ref{F-diag}(c) show snapshots of amplitude narrowband maps for 94 and 193 \r{A} at three different instants of time (spaced 10 minutes apart). The time variation of the bright and dark regions of the waves inside the loops shows a static wave behaviour. This characteristic was seen in all data cubes we studied and during all stages of the flare, which is evidence of standing waves in the ROI.

For the 9-minute wave band we saw the same behaviour, and for this we only present narrowband maps for the 19-minute waves.

We fitted the temporal signal spatially integrated over the cube formed with the sequence of narrowband maps for the 9- and 19-minute bands. To do this, we used the function $f(t)=A_0 + A \sin(2\pi/P + \delta) \exp(t/\tau)$,  where $t$, $A$, $A_0$, $P$, $\delta$, $\tau$ are time, amplitude, vertical shift, period, phase, and decay time, respectively. The error in the fitting was approximately of 5\%. This percentage was computed as $E=|Signal-Fitting|/Signal$.  In Table \ref{T-dt} we present the parameters resulting from the best-fit function for each EUV bandpass.

\begin{table}[h!]
\caption{Fitting parameters for waves in the 9- and 19-minute bands. Subscripts a and b refer to loops \textbf{a} and \textbf{b} of Figure \ref{F-rois}. }
\label{T-dt}
\begin{tabular}{lccccr} \hline
 Wavelength  & \hspace{0.6cm} $A$ ($10^5$) \hspace{0.6cm} & \hspace{0.6cm} $A_0$ ($10^2$) \hspace{0.5cm} & \hspace{0.4cm} $P$ \hspace{0.4cm} & \hspace{0.4cm} $\delta$ \hspace{0.4cm} & \hspace{0.4cm} $\tau$ \\
(\r{A}) & (arbitrary units) & (arbitrary units) & (min) & (rad) & (min) \\ \hline
\multicolumn{6}{l}{9-minute band} \\ %\hline
94a &  1.1 & ~-0.4 & 8.5 & 3.0 & ~7.5 \\
131a & 3.4 & ~18.8 & 8.5 & 3.0 & ~5.8 \\
171a & 0.9 & -42.5 & 8.8 & 1.3 & 10.2 \\
171b & 2.9 & ~-6.1 & 8.6 & 3.1 & ~8.0 \\
193a & 1.8 & ~13.9 & 8.2 & 3.4 & ~7.2 \\
193b & 4.6 & ~77.6 & 8.9 & 3.1 & 10.8 \\
211a & 0.5 & ~23.6 & 8.7 & 3.2 & ~6.5 \\
211b & 5.0 & 132.3 & 8.9 & 3.1 & 10.2 \\
335a & 0.9 & ~22.4 & 8.9 & 3.0 & 10.0 \\
335b & 1.1 & ~23.3 & 9.1 & 3.2 & 10.8 \\ %\hline
\multicolumn{6}{l}{19-minute band} \\ %\hline
94  & 1.5 & ~~1.3   & 18.0 & 3.4 & 23.5 \\
131 & 3.6 & ~~8.4   & 17.3 & 3.5 & 20.9 \\
171 & 2.5 & ~53.1  & 18.5 & 2.5 & 14.5 \\
193 & 6.3 & 121.9 & 17.7 & 3.0 & 22.6 \\
211 & 5.0 & 130.0 & 18.5 & 2.8 & 21.2 \\
335 & 1.8 & ~39.8  & 18.2 & 3.2 & 24.1 \\ \hline 
\end{tabular}

\end{table}

When we compare Figures \ref{F-mdarc}$-$\ref{F-diag} with Table \ref{T-dt}, the fitted amplitude is two orders of magnitude higher because we did not normalize the signal that we fitted.  On the other hand, the results of fitted curves we used to search for the parameters of the 9-minute waves were different in loops \textbf{a} and \textbf{b}. However, these results were similar for the 19-minute waves in both loops.

The decay times $\tau_{9min}$ $\approx$ 5$-$10 min and $\tau_{19min} \approx$ 14$-$24 min are in agreement with values found in the literature \citep{WSCID02,WSIC03} and are characteristic of magnetoacoustic waves.

\section{Three-Dimensional Reconstruction of the Magnetic Field}
      \label{S-features}   
      
To study the wave behaviour along the loops, we modelled the magnetic field using an observed magnetogram. To do this, we extrapolated the magnetic field from the magnetogram at the photosphere to find the 3D structure of the loops to better estimate the sound and Alfv{\'e}n speeds along the field lines.  We used the force-free field approximation ($\nabla \times \vec{B}=\alpha \vec{B}$) and only considered the linear set of solutions \citep{NR72,SSC08}.  We selected a region in the LOS magnetogram with dimensions $307'' \times 307''$ that encloses the ROI. Because this region is located far from the disk centre (S19E53), we rotated it to the centre and re-sampled to obtain a regular grid before we extrapolated and rotated the field lines back. Although this is an over-simplification for the LOS magnetogram, we found magnetic field lines for $\alpha=0.02$ px$^{-1}$ that reasonably match the observed EUV loops.      
         
\begin{figure}[h!]
 \centerline{%
  \includegraphics[width=0.5\textwidth,clip=]{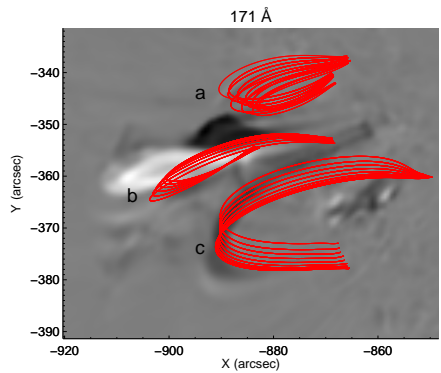}%
  }%
\caption{Magnetic field lines projected over a narrowband map in 171 \r{A} for waves in the 19-minute band in the set of loops shown in Figure \ref{F-rois}.}
\label{F-lin2d}
\end{figure}

Figure \ref{F-lin2d} shows three sets of lines extrapolated and projected over a snapshot of the dynamic map in 171 \r{A}.  Loops \textbf{a}, \textbf{b}, and \textbf{c} correspond to the loops shown in Figure \ref{F-rois}.

\subsection{Magnetic Field Lines and Velocity of the Waves} 
\label{S-lines}

We found total loop lengths of $L\approx$ 34$-$80 Mm and we considered a wavelength $\lambda=2L$ for the global standing mode.  We estimated the phase speed required as $v_{ph}=2L/P$, where $P$ is the observed period  \citep{A04}.  In Table \ref{cs-m} the loop length and phase speeds for waves in the 9- and 19-minute bands are presented in their ranges for the set of loops shown in Figure \ref{F-lin2d}. However, summing the equation $P=\sum 2\Delta x / v_{ph}$, where $\Delta x$ is the discrete displacement in the field line, we find exactly the same oscillation period as we did using our calculated velocities $c_s$ and $v_a$ (see columns 5 and 6 in Table \ref{cs-m}).

\newpage

\begin{table}[h!]
\caption{Length and phase speed for the 9- and 19-minute band required in loops of Figure \ref{F-lin2d}. $P_{c_s}$ and $P_{v_a}$ are the oscillation period calculated for $c_s$ and $v_a$ respectively.}
\label{cs-m}
\begin{tabular}{lccccr} \hline 
Loops & \hspace{0.4cm} $L $ \hspace{0.4cm} & \hspace{0.4cm} $v_{ph}$ 9-min \hspace{0.4cm} & \hspace{0.4cm} $v_{ph}$ 19-min \hspace{0.4cm} & \hspace{0.4cm} $P_{c_s}$ \hspace{0.4cm} & \hspace{0.4cm} $P_{v_a}$ \\
 		       & (Mm) & (km s$^{-1}$) & (km s$^{-1}$) & (min) & (min) \\ \hline
  a & 34$-$45 & 124.9$-$168.2 & 59.2$-$79.7  & 19.6 & 80 \\
  b & 60$-$68 & 222.3$-$252.9 & 105.3$-$119.8 & 19.6 & 28 \\
  c & 76$-$80 & 281.9$-$299.6 & 133.5$-$141.9 & 23.5 & ~92.6 \\ \hline
\end{tabular}	
\end{table}

We used two models of the solar atmosphere to estimate the ion number density $n_i$ (cm$^{-3}$) and temperature $T$ (K) as a function of loop height $h$ (km). Then we calculated the acoustic and Alfv\'{e}n speeds with the following equations \citep{A04}:

\begin{equation}
c_s=147\sqrt{T/10^6} \quad \textnormal{km s}^{-1},
\label{E-cs}
\end{equation} 
and
\begin{equation}
v_a=\frac{B}{\sqrt{4 \pi \rho}} \quad \textnormal{km s}^{-1},
\label{E-va}
\end{equation}
where $\rho \approx m_p n_i$ is the ion mass density and $m_p$ the proton mass.\\

To find $n_i(h)$, we used a chromosphere model for flare regions that satisfy $h \le 1459$ km \citep{MAVN80}. For greater heights we used a chromosphere and transition region model for the quiet Sun \citep{AL08}. For the temperature, we used the flare model in $h \le 1459$ km. For $h > 1459$ km we applied the Solar Software (SSW) routine for the automated temperature and emission measure analysis of coronal loops and active regions developed by \cite{A13}\footnote{ http://www.lmsal.com/$\sim$aschwand/software}. The $T$ we found is in agreement with the flare-like temperatures for the six AIA bandpasses \citep{OZMWT10}. Thus, we calculated the sound speeds expected for each $T$ listed in Table \ref{T-cs}.

\begin{table}[h!]
  \caption{Sound speeds according the flare-like temperature for SDO/AIA EUV bandpasses \citep{OZMWT10}.}
  \label{T-cs}
  \begin{tabular}{lcr} \hline 
 Wavelength & \hspace{1.3cm} $T$ \hspace{1.3cm} & \hspace{1.3cm} $c_s$ \hspace{0.6cm}  \\
	(\r{A}) & (MK)  & (km s$^{-1}$) \\ \hline
		131 & 11.22 & 492.39 \\
		171 & ~0.70 & 122.98 \\
		193 & 17.78 & 619.84 \\
		211 & ~1.99 & 207.36 \\
		335 & ~2.81 & 246.41 \\
		94  & ~7.07 & 390.86  \\ \hline
   \end{tabular}
  \end{table}

\newpage

\begin{figure}[h!] 
\centerline{%
\includegraphics[width=0.8\textwidth,clip=]{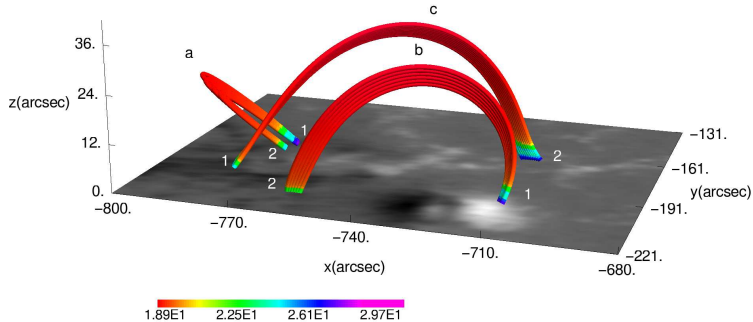}%
}%
\caption{Ion number density ($n_i$ cm$^{-3}$) along loops \textbf{a}, \textbf{b}, and \textbf{c}. Foot points 1 and 2 are indicated for all loops.}
\label{F-den3d}
\end{figure}

According the height $h$ (km), we interpolated $n_i$ and $T$ to calculate $c_s$ and $v_a$ for each point of the line. The ion number density distribution along the field lines is shown in Figure \ref{F-den3d}. This was plotted over the magnetogram moved to centre of solar disk. The number density varies from $n_i \approx 10^{17}$ cm$^{-3}$ (loop foot point) to $n_i \approx 10^{8}$ cm$^{-3}$ (loop apex).

For $h=0$ km we found $n_i\approx 10^{17}$ cm$^{-3}$, $T \approx 10^3$ K and  $c_s \approx 11$ km s$^{-1}$. Because the magnetic field is not uniform along the loops, we show the difference between magnetic field and Alfv{\'e}n speed values at the foot points in Table \ref{T-mresult} and also the height where $v_a$ reaches the maximum value in Table \ref{T-mapex}.

For loops \textbf{a}, \textbf{b}, and \textbf{c} we found that $B_2 < B_1$, $v_{a_2} < v_{a_1} $ and $v_a \gg c_s$ (see Table \ref{T-mresult}). The maximum value of Alfv{\'e}n speed was not in the loop apex, but at a lower height where  $B \approx$ 100$-$300 G and $n_i \approx 10^8$ cm$^{-3}$ (see Table \ref{T-mapex}). These speeds are compatible with the Alfv{\'e}n speed model in the literature, \textit{e.g.} \cite{BWHS92,SHBG92,SHBW94}.

\begin{table}[h!]
 \caption{Magnetic field and the Alfv{\'e}n and sound speeds at the foot points of loops \textbf{a}, \textbf{b}, and \textbf{c}. 
$B_1$, $B_2$, $v_{a_1}$, and $v_{a_2}$ are the magnetic field and the Alfv{\'e}n speed at foot points 1 and 2, respectively (see Figure \ref{F-va3d}).}
 \label{T-mresult}
  \begin{tabular}{lcccccr}\hline
  Loops & \hspace{0.4cm} $B_1$ \hspace{0.4cm} & \hspace{0.4cm} $B_2$ \hspace{0.4cm} & \hspace{0.4cm} $v_{a_1}$ \hspace{0.4cm} & \hspace{0.4cm} $v_{a_2}$ \hspace{0.4cm} & \hspace{0.4cm} $c_s/v_{a_1}$ \hspace{0.2cm} & \hspace{0.2cm} $c_s/v_{a_2}$ \\
		&  (G) & (G) & (km s$^{-1}$) & (km s$^{-1}$) & & \\ \hline 
  	  a & 144.84 & ~91.88 & 0.97 & 0.59 & 11.41 & 18.80 \\
	  b & 721.18 & 125.98 & 8.89 & 0.81 & ~1.17 & 13.71 \\
	  c & 268.54 & ~48.70 & 3.38 & 0.31 & ~3.06 & 35.47 \\ \hline
   \end{tabular}		
\end{table}
\begin{table}[h!]
 \caption{Height where the Alfv{\'e}n speed reached its maximum value in loops \textbf{a}, \textbf{b}, and \textbf{c} and corresponding physical parameters.}
  \label{T-mapex}
  \begin{tabular}{lccccccr} \hline 
  Loops & $h$ & $n_e$ & $T$ & $B$ & $c_s$ & $v_a$ & $c_s/v_a$\\
	& (km) & (cm$^{-3}$) & (K) & (G) & (km s$^{-1}$) & (km s$^{-1}$) & \\ \hline
  \textbf{a} & 8254.60 & $3.32\times 10^8$ & $3.23\times 10^6$ & 102.05 & 264.57 &  12212.30 & 0.020 \\
  \textbf{b} & 2820.19 & $8.27\times 10^8$ & $1.06\times 10^6$ & 299.40 & 151.46 &  22706.10 & 0.006 \\
  \textbf{c} & 4254.52 & $5.24\times 10^8$ & $2.80\times 10^6$ & 132.62 & 250.24 &  12634.50 & 0.010 \\ \hline 
  \end{tabular}
\end{table}

\begin{figure}[h!]
\centerline{%
 \includegraphics[width=0.8\textwidth,clip=]{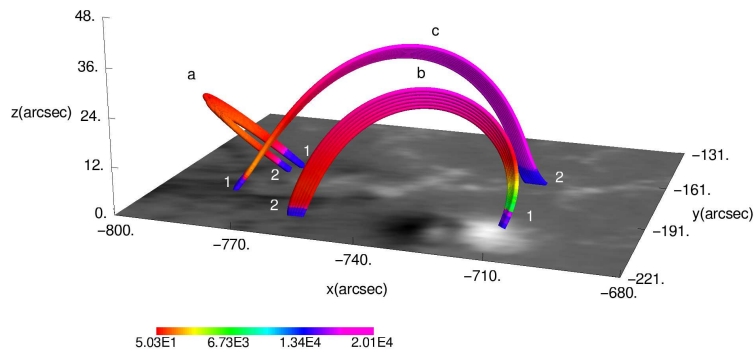}%
}%
\caption{Alfv\'{e}n speed ($v_a$ km s$^{-1}$) along loops \textbf{a}, \textbf{b}, and \textbf{c}. Foot points 1 and 2 are indicated for all loops.}
\label{F-va3d}
\end{figure}

\begin{figure}[h!] 
\centerline{%
 \includegraphics[width=0.8\textwidth,clip=]{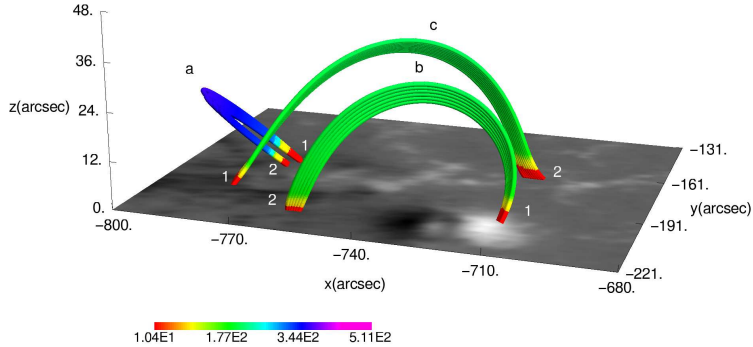}%
}%
\caption{Sound speed ($c_s$ km s$^{-1}$) along loops \textbf{a}, \textbf{b}, and \textbf{c}. Foot points 1 and 2 are indicated for all loops.}
\label{F-cs3d}
\end{figure}

In Figure \ref{F-va3d} we show the distribution of Alfv{\'e}n speeds along the loops. In Figure \ref{F-cs3d} we show the same for the sound speed. The maximum value of the temperature and sound speed for loop \textbf{a} is consistent with the fact that the flare started in this loop. Observing the loops in Figures \ref{F-mdarc} and \ref{F-mdrb}, we see that the brightness distribution along the loops resembles the Alfv{\'e}n speed variation in Figure \ref{F-va3d}. This is due to the asymmetric variation of magnetic field. This is most clear for loop \textbf{c} in the dynamical narrowband maps where the amplitude modulation is stronger in one leg, starting at the northern foot point (see Figure \ref{F-mdrb}).

Although the Alfv{\'e}n speed distribution in space is compatible with the brightness distribution of the observed waves, the calculated periods in Table \ref{cs-m} are not compatible with the observed periods. Unfortunately, the periods obtained using the Alfv{\'e}n speed are too long. The reason might be that the ion density we used to calculate $v_a$ is lower. To obtain periods that are closer to 19 minutes, it would be more convenient to use density values sixteen times higher. When the sound speed distribution is used, the period of 19 minutes is compatible. However, the loops show a symmetrical brightness distribution.

These waves were present from before the onset until the end of a compact-type flare with a modulation depth of 20\%$-$60\% for 131, 335, and 94 \r{A} and $\approx$8\% for 171, 193, and 211 \r{A}. We may relate these waves to QPP of long periods, as suggested by the mechanism based upon MHD oscillations \citep{NM09}, for example.

\section{Summary} 
\label{S-otro}

We analysed Active Region 11272 where a C1.9 flare occurred on SOL2011-08-17.  We found standing waves with periods in the 9- and 19-minute bands. The waves were detected using the PWF method \citep{SN08,SNAO10}. The dynamical narrowband maps showed clear standing waves. The 19-minute waves were detected in six EUV bandpasses during one hour, with decay time $\tau \sim$ 14$-$24 minutes in loops \textbf{a}, \textbf{b}, and \textbf{c}. Nine-minute waves were detected in six EUV bands in loop \textbf{a}. For loop \textbf{b} they were detected in all bands except for 131 and 94 \r{A}. Loop \textbf{c} did not show this period. The decay time for these waves was $\tau \sim$ 5$-$10 minutes. It is important to note that the spatial distribution of the waves is spatially coincident with the observed loops in AIA images before we applied the PWF method. This was observed in the dynamical narrowband maps that were constructed for the amplitude (see snapshots in Figures \ref{F-mdarc}$-$\ref{F-diag}).

We studied the wave behaviour along the loops using the field lines that we extrapolated over the LOS magnetogram. We calculated the sound speed using the temperature values obtained from the atmosphere models for flare regions in $h< 1459$ km \citep{MAVN80} and for coronal heights we used the routines of SSW for automated temperature analysis \citep{A13}. The $c_s$ calculated were compared with the phase speed expected for the length of the loops in periods of 9 and 19 minutes. 

We calculated the Alfv{\'e}n speed using the magnetic field value obtained from the extrapolation and the ion number density from two solar atmosphere models. Flare model for $h<1459$ km \citep{MAVN80} and the quiet Sun model for greater heights \citep{AL08}. The magnetic field was not uniform along the loops. This asymmetry in the magnetic field of the loops is reflected in the behaviour of $v_a$, which is shown in Figure \ref{F-mdrb}. In some points of the loops $v_a \approx 10^4$ km s$^{-1}$ with $B \approx$ 100$-$340 G, but with $n_i \approx 10^8$ cm$^{-3}$. This result is in agreement with the literature \citep{BWHS92,SHBG92,SHBW94}. 

Although there is a preference for the phase velocity by Alfv{\'e}n waves to explain the observed brightness waveform, the periods are much longer than observed. We may match the observed period if we increase the densities in the atmosphere model we used, for example, through chromospheric evaporation (\textit{e.g.} \citep{A04}). 

The modulation depths for 131, 335, and 94 \r{A} are inside $\approx$ 20\%$-$60\%, and in 171, 193, and 211 \r{A} they are about 8\%. We considered the waves we analysed in this work as standing waves. They may also be associated with QPP of long periods because the modulation is detected in all EUV bandpasses.

 \begin{acks}[Acknowledgements]
 	We would like to thank Brazilian agency CAPES for the financial support. We thank the anonymous referee for the careful reading and valuable suggestions.
 \end{acks}

%%% BIBLIOGRAPHY %%%%%%%%%%%%%%%%%%%%%%%%%%%%%%%%%%%%%%%%%%%%%%%%%%%%%%%%%%%

%     % format of references provided by the journal (.bst)
%\bibliographystyle{spr-mp-sola}
%     % name your Bibtex file containing your references (.bib)
%\bibliography{sola_bibliography_example}  
%\bibliography{referencias}  

\begin{thebibliography}{31}
	% BibTex style file: spr-mp-sola.bst (nameyear), 2015-03-09
	\ifx\bisbn     \undefined \def\bisbn  #1{ISBN #1}\fi
	\ifx\binits    \undefined \def\binits#1{#1}\fi
	\ifx\bauthor   \undefined \def\bauthor#1{#1}\fi
	\ifx\batitle   \undefined \def\batitle#1{#1}\fi
	\ifx\bjtitle   \undefined \def\bjtitle#1{\textit{#1}}\fi
	\ifx\bvolume   \undefined \def\bvolume#1{\textbf{#1}}\fi
	\ifx\byear     \undefined \def\byear#1{#1}\fi
	\ifx\bissue    \undefined \def\bissue#1{#1}\fi
	\ifx\bfpage    \undefined \def\bfpage#1{#1}\fi
	\ifx\blpage    \undefined \def\blpage #1{#1}\fi
	\ifx\burl      \undefined \def\burl#1{\textsf{#1}}\fi
	\ifx\href      \undefined \def\href#1#2{\textsf{#2}}\fi
	\ifx\betal     \undefined \def\betal{\textit{et al.}}\fi
	\ifx\bctitle   \undefined \def\bctitle#1{#1}\fi
	\ifx\beditor   \undefined \def\beditor#1{#1}\fi
	\ifx\bbtitle   \undefined \def\bbtitle#1{\textit{#1}}\fi
	\ifx\bedition  \undefined \def\bedition#1{#1}\fi
	\ifx\bseriesno \undefined \def\bseriesno#1{\textbf{#1}}\fi
	\ifx\blocation \undefined \def\blocation#1{#1}\fi
	\ifx\bsertitle \undefined \def\bsertitle#1{\textit{#1}}\fi
	\ifx\bsnm      \undefined \def\bsnm#1{#1}\fi
	\ifx\bsuffix   \undefined \def\bsuffix#1{#1}\fi
	\ifx\bparticle \undefined \def\bparticle#1{#1}\fi
	\ifx\barticle  \undefined \def\barticle#1{}\fi
	\ifx\binstitute  \undefined \def\binstitute#1{#1}\fi
	\ifx\bpublisher  \undefined \def\bpublisher#1{#1}\fi
	\ifx\doiurl    \undefined
	\def\doiurl#1{\href{http://dx.doi.org/#1}{\textsf{DOI}}}\fi
	\ifx\arxivurl  \undefined
	\def\arxivurl#1{\href{http://arxiv.org/abs/#1}{\textsf{arXiv}}}\fi
	\ifx\adsurl    \undefined
	\def\adsurl#1{\href{http://adsabs.harvard.edu/abs/#1}{\textsf{ADS}}}\fi
	\ifx\botherref \undefined \def\botherref#1{}\fi
	\ifx\url       \undefined \def\url#1{\textsf{#1}}\fi
	\ifx\bchapter  \undefined \def\bchapter#1{}\fi
	\ifx\bbook     \undefined \def\bbook#1{}\fi
	\ifx\bcomment  \undefined \def\bcomment#1{#1}\fi
	\ifx\oauthor   \undefined \def\oauthor#1{#1}\fi
	\ifx\citeauthoryear \undefined\def \citeauthoryear#1{#1}\fi
	\ifx\endbibitem\undefined \def\endbibitem{}\fi
	\ifx\bconflocation  \undefined \def\bconflocation#1{#1} \fi
	
	\bibitem[\protect\citeauthoryear{{Aschwanden}}{2004}]{A04}
	\begin{bbook}
		\bauthor{\bsnm{{Aschwanden}}, \binits{M.J.}}:
		\byear{2004},
		\bbtitle{{Physics of the Solar Corona. An Introduction}},
		\bpublisher{Praxis},
		\blocation{Chichester}.
		\adsurl{2004psci.book.....A}.
	\end{bbook}
	%\endbibitem
	
	\bibitem[\protect\citeauthoryear{{Aschwanden} \textit{et~al.}}{2013}]{A13}
	\begin{barticle}
		\bauthor{\bsnm{{Aschwanden}}, \binits{M.J.}},
		\bauthor{\bsnm{{Boerner}}, \binits{P.}},
		\bauthor{\bsnm{{Schrijver}}, \binits{C.J.}},
		\bauthor{\bsnm{{Malanushenko}}, \binits{A.}}:
		\byear{2013},
		\batitle{{Automated Temperature and Emission Measure Analysis of Coronal Loops
				and Active Regions Observed with the Atmospheric Imaging Assembly on the
				Solar Dynamics Observatory (SDO/AIA)}}.
		\bjtitle{\solphys}
		\bvolume{283},
		\bfpage{5}.
		\doiurl{10.1007/s11207-011-9876-5}.
		\adsurl{2013SoPh..283....5A}.
	\end{barticle}
	%\endbibitem
	
	\bibitem[\protect\citeauthoryear{{Avrett} and {Loeser}}{2008}]{AL08}
	\begin{barticle}
		\bauthor{\bsnm{{Avrett}}, \binits{E.H.}},
		\bauthor{\bsnm{{Loeser}}, \binits{R.}}:
		\byear{2008},
		\batitle{{Models of the Solar Chromosphere and Transition Region from SUMER and
				HRTS Observations: Formation of the Extreme-Ultraviolet Spectrum of Hydrogen,
				Carbon, and Oxygen}}.
		\bjtitle{\apjs}
		\bvolume{175},
		\bfpage{229}.
		\doiurl{10.1086/523671}.
		\adsurl{2008ApJS..175..229A}.
	\end{barticle}
	%\endbibitem
	
	\bibitem[\protect\citeauthoryear{{Brosius} \textit{et~al.}}{1992}]{BWHS92}
	\begin{barticle}
		\bauthor{\bsnm{{Brosius}}, \binits{J.W.}},
		\bauthor{\bsnm{{Willson}}, \binits{R.F.}},
		\bauthor{\bsnm{{Holman}}, \binits{G.D.}},
		\bauthor{\bsnm{{Schmelz}}, \binits{J.T.}}:
		\byear{1992},
		\batitle{{Coronal magnetic structures observing campaign. IV: Multiwaveband
				observations of sunspot and plage-associated coronal emission}}.
		\bjtitle{\apj}
		\bvolume{386},
		\bfpage{347}.
		\doiurl{10.1086/171021}.
		\adsurl{1992ApJ...386..347B}.
	\end{barticle}
	%\endbibitem
	
	\bibitem[\protect\citeauthoryear{{De Moortel}}{2006}]{D06}
	\begin{barticle}
		\bauthor{\bsnm{{De Moortel}}, \binits{I.}}:
		\byear{2006},
		\batitle{{Propagating magnetohydrodynamics waves in coronal loops}}.
		\bjtitle{Phil. Trans. Royal Soc. London Ser. A}
		\bvolume{364},
		\bfpage{461}.
		\doiurl{10.1098/rsta.2005.1710}.
		\adsurl{2006RSPTA.364..461D}.
	\end{barticle}
	%\endbibitem
	
	\bibitem[\protect\citeauthoryear{{De Moortel}, {Ireland}, and
		{Walsh}}{2000}]{DH00}
	\begin{barticle}
		\bauthor{\bsnm{{De Moortel}}, \binits{I.}},
		\bauthor{\bsnm{{Ireland}}, \binits{J.}},
		\bauthor{\bsnm{{Walsh}}, \binits{R.W.}}:
		\byear{2000},
		\batitle{{Observation of oscillations in coronal loops}}.
		\bjtitle{\aap}
		\bvolume{355},
		\bfpage{L23}.
		\adsurl{2000A\%26A...355L..23D}.
	\end{barticle}
	%\endbibitem
	
	\bibitem[\protect\citeauthoryear{{De Moortel}, {Munday}, and
		{Hood}}{2004}]{DMH04}
	\begin{barticle}
		\bauthor{\bsnm{{De Moortel}}, \binits{I.}},
		\bauthor{\bsnm{{Munday}}, \binits{S.A.}},
		\bauthor{\bsnm{{Hood}}, \binits{A.W.}}:
		\byear{2004},
		\batitle{{Wavelet Analysis: the effect of varying basic wavelet parameters}}.
		\bjtitle{\solphys}
		\bvolume{222},
		\bfpage{203}.
		\doiurl{10.1023/B:SOLA.0000043578.01201.2d}.
		\adsurl{2004SoPh..222..203D}.
	\end{barticle}
	%\endbibitem
	
	\bibitem[\protect\citeauthoryear{Erd{\'{e}}lyi and
		Mendoza-Brice{\~{n}}o}{2004}]{EM04}
	\begin{bchapter}
		\bauthor{\bsnm{Erd{\'{e}}lyi}, \binits{R.}},
		\bauthor{\bsnm{Mendoza-Brice{\~{n}}o}, \binits{C.A.}}:
		\byear{2004},
		\bctitle{Damping of loop oscillations in the stratified corona}.
		In: \beditor{\bsnm{Lacoste}, \binits{H.}} (ed.)
		\bbtitle{SOHO 13 Waves, Oscillations and Small-Scale Transients Events in the
			Solar Atmosphere: Joint View from SOHO and TRACE},
		\bsertitle{ESA Special Publication}
		\bseriesno{547},
		\bfpage{441}.
		\adsurl{2004ESASP.547..441E}.
	\end{bchapter}
	%\endbibitem
	
	\bibitem[\protect\citeauthoryear{Kim, Nakariakov, and Shibasaki}{2012}]{KNS12}
	\begin{barticle}
		\bauthor{\bsnm{Kim}, \binits{S.}},
		\bauthor{\bsnm{Nakariakov}, \binits{V.M.}},
		\bauthor{\bsnm{Shibasaki}, \binits{K.}}:
		\byear{2012},
		\batitle{Slow magnetoacoustic oscillations in the microwave emission of solar
			flares}.
		\bjtitle{\apjl}
		\bvolume{756},
		\bfpage{L36}.
		\doiurl{10.1088/2041-8205/756/2/L36}.
		\adsurl{2012ApJ...756L..36K}.
	\end{barticle}
	%\endbibitem
	
	\bibitem[\protect\citeauthoryear{{Kupriyanova} \textit{et~al.}}{2010}]{KMNS10}
	\begin{barticle}
		\bauthor{\bsnm{{Kupriyanova}}, \binits{E.G.}},
		\bauthor{\bsnm{{Melnikov}}, \binits{V.F.}},
		\bauthor{\bsnm{{Nakariakov}}, \binits{V.M.}},
		\bauthor{\bsnm{{Shibasaki}}, \binits{K.}}:
		\byear{2010},
		\batitle{{Types of Microwave Quasi-Periodic Pulsations in Single Flaring
				Loops}}.
		\bjtitle{\solphys}
		\bvolume{267},
		\bfpage{329}.
		\doiurl{10.1007/s11207-010-9642-0}.
		\adsurl{2010SoPh..267..329K}.
	\end{barticle}
	%\endbibitem
	
	\bibitem[\protect\citeauthoryear{{Lemen} \textit{et~al.}}{2012}]{LAB12}
	\begin{barticle}
		\bauthor{\bsnm{{Lemen}}, \binits{J.R.}},
		\bauthor{\bsnm{{Title}}, \binits{A.M.}},
		\bauthor{\bsnm{{Akin}}, \binits{D.J.}},
		\bauthor{\bsnm{{Boerner}}, \binits{P.F.}},
		\bauthor{\bsnm{{Chou}}, \binits{C.}},
		\bauthor{\bsnm{{Drake}}, \binits{J.F.}},
		\bauthor{\bsnm{{Duncan}}, \binits{D.W.}},
		\bauthor{\bsnm{{Edwards}}, \binits{C.G.}},
		\bauthor{\bsnm{{Friedlaender}}, \binits{F.M.}},
		\bauthor{\bsnm{{Heyman}}, \binits{G.F.}},
		\bauthor{\bsnm{{Hurlburt}}, \binits{N.E.}},
		\bauthor{\bsnm{{Katz}}, \binits{N.L.}},
		\bauthor{\bsnm{{Kushner}}, \binits{G.D.}},
		\bauthor{\bsnm{{Levay}}, \binits{M.}},
		\bauthor{\bsnm{{Lindgren}}, \binits{R.W.}},
		\bauthor{\bsnm{{Mathur}}, \binits{D.P.}},
		\bauthor{\bsnm{{McFeaters}}, \binits{E.L.}},
		\bauthor{\bsnm{{Mitchell}}, \binits{S.}},
		\bauthor{\bsnm{{Rehse}}, \binits{R.A.}},
		\bauthor{\bsnm{{Schrijver}}, \binits{C.J.}},
		\bauthor{\bsnm{{Springer}}, \binits{L.A.}},
		\bauthor{\bsnm{{Stern}}, \binits{R.A.}},
		\bauthor{\bsnm{{Tarbell}}, \binits{T.D.}},
		\bauthor{\bsnm{{Wuelser}}, \binits{J.-P.}},
		\bauthor{\bsnm{{Wolfson}}, \binits{C.J.}},
		\bauthor{\bsnm{{Yanari}}, \binits{C.}},
		\bauthor{\bsnm{{Bookbinder}}, \binits{J.A.}},
		\bauthor{\bsnm{{Cheimets}}, \binits{P.N.}},
		\bauthor{\bsnm{{Caldwell}}, \binits{D.}},
		\bauthor{\bsnm{{Deluca}}, \binits{E.E.}},
		\bauthor{\bsnm{{Gates}}, \binits{R.}},
		\bauthor{\bsnm{{Golub}}, \binits{L.}},
		\bauthor{\bsnm{{Park}}, \binits{S.}},
		\bauthor{\bsnm{{Podgorski}}, \binits{W.A.}},
		\bauthor{\bsnm{{Bush}}, \binits{R.I.}},
		\bauthor{\bsnm{{Scherrer}}, \binits{P.H.}},
		\bauthor{\bsnm{{Gummin}}, \binits{M.A.}},
		\bauthor{\bsnm{{Smith}}, \binits{P.}},
		\bauthor{\bsnm{{Auker}}, \binits{G.}},
		\bauthor{\bsnm{{Jerram}}, \binits{P.}},
		\bauthor{\bsnm{{Pool}}, \binits{P.}},
		\bauthor{\bsnm{{Soufli}}, \binits{R.}},
		\bauthor{\bsnm{{Windt}}, \binits{D.L.}},
		\bauthor{\bsnm{{Beardsley}}, \binits{S.}},
		\bauthor{\bsnm{{Clapp}}, \binits{M.}},
		\bauthor{\bsnm{{Lang}}, \binits{J.}},
		\bauthor{\bsnm{{Waltham}}, \binits{N.}}:
		\byear{2012},
		\batitle{{The Atmospheric Imaging Assembly (AIA) on the Solar Dynamics
				Observatory (SDO)}}.
		\bjtitle{\solphys}
		\bvolume{275},
		\bfpage{17}.
		\doiurl{10.1007/s11207-011-9776-8}.
		\adsurl{2012SoPh..275...17L}.
	\end{barticle}
	%\endbibitem
	
	\bibitem[\protect\citeauthoryear{{Leroy} and {Schwartz}}{1982}]{LS82}
	\begin{barticle}
		\bauthor{\bsnm{{Leroy}}, \binits{B.}},
		\bauthor{\bsnm{{Schwartz}}, \binits{S.J.}}:
		\byear{1982},
		\batitle{{Propagation of waves in an atmosphere in the presence of a magnetic
				field. V: The theory of magneto-acoustic-gravity oscillations.}}
		\bjtitle{\aap}
		\bvolume{112},
		\bfpage{84}.
		\adsurl{1982A\%26A...112...84L}.
	\end{barticle}
	%\endbibitem
	
	\bibitem[\protect\citeauthoryear{Machado \textit{et~al.}}{1980}]{MAVN80}
	\begin{barticle}
		\bauthor{\bsnm{Machado}, \binits{M.E.}},
		\bauthor{\bsnm{Avrett}, \binits{E.H.}},
		\bauthor{\bsnm{Vernazza}, \binits{J.E.}},
		\bauthor{\bsnm{Noyes}, \binits{R.W.}}:
		\byear{1980},
		\batitle{{Semiempirical models of chromospheric flare regions}}.
		\bjtitle{\apj}
		\bvolume{242},
		\bfpage{336}.
		\doiurl{10.1086/158467}.
		\adsurl{1980ApJ...242..336M}.
	\end{barticle}
	%\endbibitem
	
	\bibitem[\protect\citeauthoryear{{Mathioudakis}, {Jess}, and
		{Erd{\'e}lyi}}{2013}]{MJE13}
	\begin{barticle}
		\bauthor{\bsnm{{Mathioudakis}}, \binits{M.}},
		\bauthor{\bsnm{{Jess}}, \binits{D.B.}},
		\bauthor{\bsnm{{Erd{\'e}lyi}}, \binits{R.}}:
		\byear{2013},
		\batitle{{Alfv{\'e}n Waves in the Solar Atmosphere. From Theory to
				Observations}}.
		\bjtitle{\ssr}
		\bvolume{175},
		\bfpage{1}.
		\doiurl{10.1007/s11214-012-9944-7}.
		\adsurl{2013SSRv..175....1M}.
	\end{barticle}
	%\endbibitem
	
	\bibitem[\protect\citeauthoryear{{Mendoza-Brice{\~n}o}, {Erd{\'e}lyi}, and
		{Sigalotti}}{2004}]{MES04}
	\begin{barticle}
		\bauthor{\bsnm{{Mendoza-Brice{\~n}o}}, \binits{C.A.}},
		\bauthor{\bsnm{{Erd{\'e}lyi}}, \binits{R.}},
		\bauthor{\bsnm{{Sigalotti}}, \binits{L.D.G.}}:
		\byear{2004},
		\batitle{{The Effects of Stratification on Oscillating Coronal Loops}}.
		\bjtitle{\apj}
		\bvolume{605},
		\bfpage{493}.
		\doiurl{10.1086/382182}.
		\adsurl{2004ApJ...605..493M}.
	\end{barticle}
	%\endbibitem
	
	\bibitem[\protect\citeauthoryear{{Morton}, {Hood}, and
		{Erd{\'e}lyi}}{2010}]{MHE10}
	\begin{barticle}
		\bauthor{\bsnm{{Morton}}, \binits{R.J.}},
		\bauthor{\bsnm{{Hood}}, \binits{A.W.}},
		\bauthor{\bsnm{{Erd{\'e}lyi}}, \binits{R.}}:
		\byear{2010},
		\batitle{{Propagating magneto-hydrodynamic waves in a cooling homogenous
				coronal plasma}}.
		\bjtitle{\aap}
		\bvolume{512},
		\bfpage{A23}.
		\doiurl{10.1051/0004-6361/200913365}.
		\adsurl{2010A\%26A...512A..23M}.
	\end{barticle}
	%\endbibitem
	
	\bibitem[\protect\citeauthoryear{{Nakagawa} and {Raadu}}{1972}]{NR72}
	\begin{barticle}
		\bauthor{\bsnm{{Nakagawa}}, \binits{Y.}},
		\bauthor{\bsnm{{Raadu}}, \binits{M.A.}}:
		\byear{1972},
		\batitle{{On Practical Representation of Magnetic Field}}.
		\bjtitle{\solphys}
		\bvolume{25},
		\bfpage{127}.
		\doiurl{10.1007/BF00155751}.
		\adsurl{1972SoPh...25..127N}.
	\end{barticle}
	%\endbibitem
	
	\bibitem[\protect\citeauthoryear{{Nakariakov} and {Melnikov}}{2009}]{NM09}
	\begin{barticle}
		\bauthor{\bsnm{{Nakariakov}}, \binits{V.M.}},
		\bauthor{\bsnm{{Melnikov}}, \binits{V.F.}}:
		\byear{2009},
		\batitle{{Quasi-Periodic Pulsations in Solar Flares}}.
		\bjtitle{\ssr}
		\bvolume{149},
		\bfpage{119}.
		\doiurl{10.1007/s11214-009-9536-3}.
		\adsurl{2009SSRv..149..119N}.
	\end{barticle}
	%\endbibitem
	
	\bibitem[\protect\citeauthoryear{{Nistic{\`o}}, {Nakariakov}, and
		{Verwichte}}{2013}]{NNV13}
	\begin{barticle}
		\bauthor{\bsnm{{Nistic{\`o}}}, \binits{G.}},
		\bauthor{\bsnm{{Nakariakov}}, \binits{V.M.}},
		\bauthor{\bsnm{{Verwichte}}, \binits{E.}}:
		\byear{2013},
		\batitle{{Decaying and decayless transverse oscillations of a coronal loop}}.
		\bjtitle{\aap}
		\bvolume{552},
		\bfpage{A57}.
		\doiurl{10.1051/0004-6361/201220676}.
		\adsurl{2013A\%26A...552A..57N}.
	\end{barticle}
	%\endbibitem
	
	\bibitem[\protect\citeauthoryear{{O'Dwyer} \textit{et~al.}}{2010}]{OZMWT10}
	\begin{barticle}
		\bauthor{\bsnm{{O'Dwyer}}, \binits{B.}},
		\bauthor{\bsnm{{Del Zanna}}, \binits{G.}},
		\bauthor{\bsnm{{Mason}}, \binits{H.E.}},
		\bauthor{\bsnm{{Weber}}, \binits{M.A.}},
		\bauthor{\bsnm{{Tripathi}}, \binits{D.}}:
		\byear{2010},
		\batitle{{SDO/AIA response to coronal hole, quiet Sun, active region, and flare
				plasma}}.
		\bjtitle{\aap}
		\bvolume{521},
		\bfpage{A21}.
		\doiurl{10.1051/0004-6361/201014872}.
		\adsurl{2010A\%26A...521A..21O}.
	\end{barticle}
	%\endbibitem
	
	\bibitem[\protect\citeauthoryear{{Roberts}}{1981a}]{R81a}
	\begin{barticle}
		\bauthor{\bsnm{{Roberts}}, \binits{B.}}:
		\byear{1981}a,
		\batitle{{Wave propagation in a magnetically structured atmosphere. I: Surface
				waves at a magnetic interface.}}
		\bjtitle{\solphys}
		\bvolume{69},
		\bfpage{27}.
		\doiurl{10.1007/BF00151253}.
		\adsurl{1981SoPh...69...27R}.
	\end{barticle}
	%\endbibitem
	
	\bibitem[\protect\citeauthoryear{{Roberts}}{1981b}]{R81b}
	\begin{barticle}
		\bauthor{\bsnm{{Roberts}}, \binits{B.}}:
		\byear{1981}b,
		\batitle{{Wave Propagation in a Magnetically Structured Atmosphere. II: Waves
				in a Magnetic Slab}}.
		\bjtitle{\solphys}
		\bvolume{69},
		\bfpage{39}.
		\doiurl{10.1007/BF00151254}.
		\adsurl{1981SoPh...69...39R}.
	\end{barticle}
	%\endbibitem
	
	\bibitem[\protect\citeauthoryear{{Scherrer} \textit{et~al.}}{2012}]{SSBK12}
	\begin{barticle}
		\bauthor{\bsnm{{Scherrer}}, \binits{P.H.}},
		\bauthor{\bsnm{{Schou}}, \binits{J.}},
		\bauthor{\bsnm{{Bush}}, \binits{R.I.}},
		\bauthor{\bsnm{{Kosovichev}}, \binits{A.G.}},
		\bauthor{\bsnm{{Bogart}}, \binits{R.S.}},
		\bauthor{\bsnm{{Hoeksema}}, \binits{J.T.}},
		\bauthor{\bsnm{{Liu}}, \binits{Y.}},
		\bauthor{\bsnm{{Duvall}}, \binits{T.L.}},
		\bauthor{\bsnm{{Zhao}}, \binits{J.}},
		\bauthor{\bsnm{{Title}}, \binits{A.M.}},
		\bauthor{\bsnm{{Schrijver}}, \binits{C.J.}},
		\bauthor{\bsnm{{Tarbell}}, \binits{T.D.}},
		\bauthor{\bsnm{{Tomczyk}}, \binits{S.}}:
		\byear{2012},
		\batitle{{The Helioseismic and Magnetic Imager (HMI) Investigation for the
				Solar Dynamics Observatory (SDO)}}.
		\bjtitle{\solphys}
		\bvolume{275},
		\bfpage{207}.
		\doiurl{10.1007/s11207-011-9834-2}.
		\adsurl{2012SoPh..275..207S}.
	\end{barticle}
	%\endbibitem
	
	\bibitem[\protect\citeauthoryear{{Schmelz} \textit{et~al.}}{1992}]{SHBG92}
	\begin{barticle}
		\bauthor{\bsnm{{Schmelz}}, \binits{J.T.}},
		\bauthor{\bsnm{{Holman}}, \binits{G.D.}},
		\bauthor{\bsnm{{Brosius}}, \binits{J.W.}},
		\bauthor{\bsnm{{Gonzalez}}, \binits{R.D.}}:
		\byear{1992},
		\batitle{{Coronal Magnetic Structures Observing Campaign. II: Magnetic and
				plasma properties of a solar active region}}.
		\bjtitle{\apj}
		\bvolume{399},
		\bfpage{733}.
		\doiurl{10.1086/171966}.
		\adsurl{1992ApJ...399..733S}.
	\end{barticle}
	%\endbibitem
	
	\bibitem[\protect\citeauthoryear{{Schmelz} \textit{et~al.}}{1994}]{SHBW94}
	\begin{barticle}
		\bauthor{\bsnm{{Schmelz}}, \binits{J.T.}},
		\bauthor{\bsnm{{Holman}}, \binits{G.D.}},
		\bauthor{\bsnm{{Brosius}}, \binits{J.W.}},
		\bauthor{\bsnm{{Willson}}, \binits{R.F.}}:
		\byear{1994},
		\batitle{{Coronal Magnetic Structures Observing Campaign. III: Coronal plasma
				and magnetic field diagnostics derived from multiwaveband active region
				observations}}.
		\bjtitle{\apj}
		\bvolume{434},
		\bfpage{786}.
		\doiurl{10.1086/174781}.
		\adsurl{1994ApJ...434..786S}.
	\end{barticle}
	%\endbibitem
	
	\bibitem[\protect\citeauthoryear{{Selhorst}, {Silva-V{\'a}lio}, and
		{Costa}}{2008}]{SSC08}
	\begin{barticle}
		\bauthor{\bsnm{{Selhorst}}, \binits{C.L.}},
		\bauthor{\bsnm{{Silva-V{\'a}lio}}, \binits{A.}},
		\bauthor{\bsnm{{Costa}}, \binits{J.E.R.}}:
		\byear{2008},
		\batitle{{Solar atmospheric model over a highly polarized 17 GHz active
				region}}.
		\bjtitle{\aap}
		\bvolume{488},
		\bfpage{1079}.
		\doiurl{10.1051/0004-6361:20079217}.
		\adsurl{2008A\%26A...488.1079S}.
	\end{barticle}
	%\endbibitem
	
	\bibitem[\protect\citeauthoryear{{Sych} and {Nakariakov}}{2008}]{SN08}
	\begin{barticle}
		\bauthor{\bsnm{{Sych}}, \binits{R.A.}},
		\bauthor{\bsnm{{Nakariakov}}, \binits{V.M.}}:
		\byear{2008},
		\batitle{{The Pixelised Wavelet Filtering Method to Study Waves and
				Oscillations in Time Sequences of Solar Atmospheric Images}}.
		\bjtitle{\solphys}
		\bvolume{248},
		\bfpage{395}.
		\doiurl{10.1007/s11207-007-9005-7}.
		\adsurl{2008SoPh..248..395S}.
	\end{barticle}
	%\endbibitem
	
	\bibitem[\protect\citeauthoryear{{Sych} \textit{et~al.}}{2010}]{SNAO10}
	\begin{barticle}
		\bauthor{\bsnm{{Sych}}, \binits{R.A.}},
		\bauthor{\bsnm{{Nakariakov}}, \binits{V.M.}},
		\bauthor{\bsnm{{Anfinogentov}}, \binits{S.A.}},
		\bauthor{\bsnm{{Ofman}}, \binits{L.}}:
		\byear{2010},
		\batitle{{Web-Based Data Processing System for Automated Detection of
				Oscillations with Applications to the Solar Atmosphere}}.
		\bjtitle{\solphys}
		\bvolume{266},
		\bfpage{349}.
		\doiurl{10.1007/s11207-010-9616-2}.
		\adsurl{2010SoPh..266..349S}.
	\end{barticle}
	%\endbibitem
	
	\bibitem[\protect\citeauthoryear{{Torrence} and {Compo}}{1998}]{TC98}
	\begin{barticle}
		\bauthor{\bsnm{{Torrence}}, \binits{C.}},
		\bauthor{\bsnm{{Compo}}, \binits{G.P.}}:
		\byear{1998},
		\batitle{{A Practical Guide to Wavelet Analysis.}}
		\bjtitle{Bulletin of the American Meteorological Society}
		\bvolume{79},
		\bfpage{61}.
		\doiurl{10.1175/1520-0477(1998)079<0061:APGTWA>2.0.CO;2}.
		\adsurl{1998BAMS...79...61T}.
	\end{barticle}
	%\endbibitem
	
	\bibitem[\protect\citeauthoryear{{Wang} \textit{et~al.}}{2003}]{WSIC03}
	\begin{barticle}
		\bauthor{\bsnm{{Wang}}, \binits{T.J.}},
		\bauthor{\bsnm{{Solanki}}, \binits{S.K.}},
		\bauthor{\bsnm{{Innes}}, \binits{D.E.}},
		\bauthor{\bsnm{{Curdt}}, \binits{W.}},
		\bauthor{\bsnm{{Marsch}}, \binits{E.}}:
		\byear{2003},
		\batitle{{Slow-mode standing waves observed by SUMER in hot coronal loops}}.
		\bjtitle{\aap}
		\bvolume{402},
		\bfpage{L17}.
		\doiurl{10.1051/0004-6361:20030448}.
		\adsurl{2003A\%26A...402L..17W}.
	\end{barticle}
	%\endbibitem
	
	\bibitem[\protect\citeauthoryear{{Wang} \textit{et~al.}}{2002}]{WSCID02}
	\begin{barticle}
		\bauthor{\bsnm{{Wang}}, \binits{T.}},
		\bauthor{\bsnm{{Solanki}}, \binits{S.K.}},
		\bauthor{\bsnm{{Curdt}}, \binits{W.}},
		\bauthor{\bsnm{{Innes}}, \binits{D.E.}},
		\bauthor{\bsnm{{Dammasch}}, \binits{I.E.}}:
		\byear{2002},
		\batitle{{Doppler Shift Oscillations of Hot Solar Coronal Plasma Seen by SUMER:
				A Signature of Loop Oscillations?}}
		\bjtitle{\apjl}
		\bvolume{574},
		\bfpage{L101}.
		\doiurl{10.1086/342189}.
		\adsurl{2002ApJ...574L.101W}.
	\end{barticle}
	%\endbibitem
	
\end{thebibliography}

\end{article} 

\end{document}